# Parametric instabilities and their control in advanced interferometer GW detectors

C. Zhao, L. Ju, J. Degallaix, S. Gras, and D. G. Blair

*School of Physics, the University of Western Australia, 35 Stirling Highway, Nedlands, WA 6009*

*A detailed simulation of Advanced LIGO test mass optical cavities shows that parametric instabilities will excite acoustic modes in the test masses in the frequency range 28-35 kHz and 64-72 kHz. Using nominal Advanced LIGO optical cavity parameters with fused silica test masses, parametric instability excites 7 acoustic modes in each test mass, with parametric gain R up to 7. For the alternative sapphire test masses only 1 acoustic mode is excited in each test mass with R ~ 2. Fine tuning of the test mass radii of curvature cause the instabilities to sweep through various modes with R as high as ~2000. Sapphire test mass cavities can be tuned to completely eliminate instabilities using thermal g-factor tuning with negligible degradation of the noise performance. In the case of fused silica test mass, instabilities can be minimized but not eliminated.*

To achieve sufficient sensitivity to detect numerous predicted sources of gravitational waves, the three long baseline laser interferometer gravitational wave detectors [1, 2, 3, 4] need to achieve about one order of magnitude improved sensitivity. This improvement is planned to be achieved using larger lower acoustic loss test masses and substantially higher laser power [5]. It has already been pointed out that this improvement brings with it the risk of parametric instability [6, 7, 8]. The instability arises due to the potential for acoustic normal modes of the test masses to scatter light from the fundamental optical cavity mode into a nearby higher order mode, mediated by the radiation pressure force of the optical modes acting on the acoustic mode. The instability can occur if two conditions are met. Firstly there must be a substantial spatial overlap of the acoustic mode shape with the higher order cavity mode shape. Secondly the optical frequency difference between the cavity fundamental mode and higher order mode must match the acoustic mode frequency.

Parametric instability was observed and controlled in the niobium bar gravitational wave detector NIOBE [9]. If not controlled, instabilities cause acoustic modes to ring at very large amplitudes, sufficient to disrupt operation of a sensitive detector.

We show here that for the proposed Advanced LIGO (AdvLIGO) parameters, the conditions for instability are indeed met for a number of acoustic modes, specifically in the frequency range 28-35kHz and 64-72 kHz. Because the Young's modulus and density are smaller for fused silica than for sapphire, the acoustic mode density of a fused silica test masse is much greater than that of a sapphire test mass with the same weight. In consequence the number of parametrically unstable modes is much greater for fused silica, and they generally have higher parametric gain. After demonstrating the magnitude of the instabilities, we will present a method by which the parametric instabilities may be detuned. Again this is more effective for sapphire than for fused silica because the mode spacing in the relevant frequency range is about 6 times greater for sapphire than fused silica. We will also show that it is unlikely to be possible to predesign against the parametric instabilities unless (a) the error of calculating normal mode frequencies in standard Finite Element Modeling (FEM) software can be improved to less than the cavity bandwidth (~30 Hz), (b) the test mass density inhomogeneity is known, (c) the mirror radius of curvature can be specified to better than 0.1%.

In an optical cavity, the frequency differences between the $TEM_{00}$ mode and $TEM_{mn}$ modes are,

$$\Delta_- = \omega_0 - \omega_1 = \frac{\pi c}{L}\left(k_1 - \frac{m+p\times n}{\pi}\arccos\left(\sqrt{(1-\frac{L}{R_1})(1-\frac{L}{R_2})}\right)\right)$$

$$\Delta_+ = \omega_{1a} - \omega_0 = \frac{\pi c}{L}\left(k_{1a} + \frac{m+p\times n}{\pi}\arccos\left(\sqrt{(1-\frac{L}{R_1})(1-\frac{L}{R_2})}\right)\right) \quad (1)$$

Here $\omega_0$ is the fundamental mode frequency, $\omega_1$ is the Stokes mode frequency and $\omega_{1a}$ is the anti-Stokes mode frequency, $L$ is the cavity length, $R_1$ and $R_2$ are the mirror radii, $k_1$ and $k_{1a}$ are longitudinal mode indices, $m$ and $n$ are transverse mode indices, and $p=1$ for the Hermite-Gaussian mode and $p=2$ for the Laguerre-Gaussian mode.

By inspection of equation 1, the fundamental modes ($m+n=0$) are symmetrically distributed around the carrier modes and the Stokes mode is compensated by an anti-Stokes mode. Higher order transverse modes ($m+n>0$) are not symmetrically distributed and do not compensate each other.

Braginsky [7] has shown that the effective parametric gain R in a power recycled interferometer is given by[*]:

$$R = \frac{2PQ_m}{mcL\omega_m^2}\left(\frac{Q_1\Lambda_1}{1+\Delta\omega_1^2/\delta_1^2} - \frac{Q_{1a}\Lambda_{1a}}{1+\Delta\omega_{1a}^2/\delta_{1a}^2}\frac{\omega_{1a}}{\omega_1}\right) \quad (2)$$

---

[*] In the unusual case that the Stokes mode is within the very narrow bandwidth ($\delta_{pr}$ [7]) of the coupled cavity of the power recycling cavity and arm cavities (~4 Hz for AdvLIGO nominal parameters), the formula of R will be different from Eq. 2 [7].

When the parametric gain exceeds unity the acoustic mode will be excited. Here $P$ is the total power inside the cavity, $Q_1$ and $Q_{1a}$ are the quality factors of the Stokes and anti-Stokes modes, $Q_m$ is the quality factor of acoustic mode, $\delta_{1(a)}=\omega_{1(a)}/2Q_{1(a)}$, m is the test mass's mass, $L$ is the cavity length, $\Delta\omega_{1(a)}=\omega_0-\omega_{1(a)}-\omega_m$ is the possible detuning from the ideal resonance case, and $\Lambda_1$ and $\Lambda_{1a}$ are the overlap factors between optical and acoustic modes. The overlap factor is defined as [6],

$$\Lambda_{1(a)} = \frac{V(\int f_0(\vec{r}_\perp) f_{1(a)}(\vec{r}_\perp) u_z d\vec{r}_\perp)^2}{\int |f_0|^2 d\vec{r}_\perp \int |f_{1(a)}|^2 d\vec{r}_\perp \int |\vec{u}|^2 dV} \quad (3)$$

Here $f_0$ and $f_{1(a)}$ describe the optical field distribution over the mirror surface for the fundamental and Stokes (anti-Stokes) modes respectively, $\vec{u}$ is the spatial displacement vector for the mechanical mode, $u_z$ is the component of $\vec{u}$ normal to the mirror surface. The integrals $\int d\vec{r}_\perp$ and $\int dV$ correspond to integration over the mirror surface and the mirror volume $V$ respectively.

Using FEM (ANSYS®) to calculate mode shapes, we have evaluated the overlap factors and calculated R for AdvLIGO test mass [10, 11] acoustic modes close to the first and the second order transverse modes. The test mass and cavity parameters used are listed in Table I and Table II. Since this letter is not a predesign of the real situation but to prove the principle the error of the FEM has not been taken into account in the simulation. This assumption will not affect the final results as it is only the frequency difference between the acoustic mode and the optical mode that affect the parametric gain, and we will tune the optical mode frequencies across a 3-5 kHz range. The simulation for acoustic modes close to cavity transverse modes higher than second order is not included, because higher order modes generally have lower parametric gain due to their diffraction losses. FIG. 1(a) shows those modes with R>1. FIG. 1(b) shows a particular acoustic mode structure with frequency (31.251 kHz) close to the frequency difference between $TEM_{00}$ and $TEM_{10}$ modes. FIG. 1(c) shows the $TEM_{10}$ mode optical field distribution. The similarity of these mode structures is apparent. The overlap factor (Eq.3) for these two modes is ~ 2.9.

Because of the large acoustic mode density there are 7 acoustic modes which have the potential (R>1) to be unstable for a fused silica test mass, compared with only 1 mode for a sapphire test mass. For the nominal AdvLIGO parameters, the maximum parametric gain R for a fused silica test mass is up to ~7, compared with ~2 for a sapphire test mass.

It is possible that the parametric gain could be much larger than the values mentioned above for several reasons. (a) Standard FEM methods for calculating the acoustic mode frequencies have errors much larger than the cavity bandwidth [7]; (b) The suspension system may change acoustic mode frequencies; (c) An error of 2m (0.1%) in the mirror radius of curvature results in a cavity mode spacing error of ~30 Hz which is equal to the cavity bandwidth. Finally thermal lensing causes the radii of curvature of test masses to vary from their nominal values. Thus the worst case, where the acoustic mode frequency is very close to the frequency difference between the fundamental mode and a high order transverse mode can not be definitely avoided. For instance, in a fused silica test mass, if the acoustic mode at frequency of 33.354 kHz (the third dot from the top in Figure 1 (a)) has a 459 Hz error; the parametric gain R could be as large as ~2000.

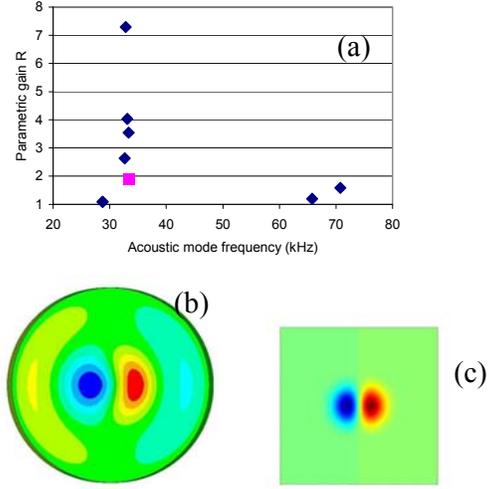

FIG. 1. (a) all the acoustic modes with R>1 for sapphire (square) and fused silica (diamond) respectively, assuming nominal AdvLIGO parameters [10, 11], (b) a typical test mass acoustic mode structure, and c) the field distribution of the cavity $TEM_{10}$ mode showing high overlap of the acoustic and optical mode structures.

**Table I. Test mass material parameters** [11]

| Parameter | Sapphire | Fused silica |
|---|---|---|
| Nominal Q | 200 million | 100 million |
| Poisson ratio | 0.23 | 0.17 |
| Young's modulus | $4\times10^{11}$ N/m$^2$ | $7\times10^{10}$ N/m$^2$ |
| Density | 3983 kg/m$^3$ | 2200 kg/m$^3$ |
| Size (Diameter × thickness) | 31.4×13 cm | 38.8×15.4 cm |
| Test mass's mass | 40 kg | 40 kg |

**Table II. Cavity parameters** [10]

| Parameter | Sapphire | Fused silica |
|---|---|---|
| Cavity length | 4 km | 4 km |
| Laser wavelength | 1064 nm | 1064 nm |
| Radius of curvature | 2.076 km | 2.076 km |
| Cavity g factor | $g_1=g_2=-0.9265$ | $g_1=g_2=-0.9265$ |
| $\delta_1/2\pi$ (m+p*n=0,1,2) | 15 Hz | 15 Hz |
| $\delta_{1a}/2\pi$ (m+p*n=6) | 191 Hz | 191 Hz |
| $\delta_{1a}/2\pi$ (m+p*n=7) | 535 Hz | 535 Hz |
| Power recycling mirror transmission | 6% | 6% |
| Laser power inside arm cavities | 830 kW | 830 kW |

The fact that the cavity mode spacing changes with the mirror radius of curvature (see Eq.1) also provides an opportunity to tune out the most unstable acoustic modes. Lawrence et al [12] and Degallaix, et al [13] have demonstrated that by using a heating ring near the front of the test mass you can adjust the test mass radius of curvature to effectively compensate for thermal lensing. The GEO project [14] has used this method to compensate the mismatch of radii of curvature of two interferometer mirrors. Here we propose a similar method, with the heating ring at the back of the test mass to tune the cavity mode frequencies. Substantial changes in radius of curvature can be achieved. FIG. 2 shows the AdvLIGO end test mass radius of curvature and the relative R,

$$\frac{R_{(\Delta T)}}{R_{(0)}} = \frac{1}{1 + (\Delta \omega_1 / \delta_1)^2} \quad (4)$$

as a function of the maximum temperature difference across the test mass when heated by a heating ring with variable heating power. Here we assume that there exists only single acoustic mode with frequency equal to the frequency difference between the fundamental mode and the first high order mode when without heating. The change of the radius of curvature from ~2.076 km to ~ 2.066 km corresponds to the maximum temperature difference across the test mass from 0 °K to ~0.11 °K for sapphire (average mirror temperature changed from 300 °K to ~302.5°K) and from 0 °K to ~1.2°K for fused silica (average mirror temperature changed from 300 °K to ~301°K). If one considers only a single acoustic mode this tuning is sufficient to reduce R to 1% of its original value. Unfortunately, there are many potential acoustic modes. When tuning the cavity modes away from a particular acoustic mode we generally increase the coupling to nearby acoustic modes. In sapphire test masses, the frequency gap is ~1 kHz. Tuning the cavity mode to a point between two acoustic modes minimizes the parametric gain of both. The acoustic mode gap of ~ 200 Hz for fused silica test masses makes such tuning much less effective. Thus we see that in fused silica (FIG. 3 (b)) it is impossible to tune to parametric gain R to less than 2. In sapphire (FIG. 3(a)) it is possible to tune the cavity away from the instabilities (R<1) at the radius of curvature around 2.092 km. FIG. 4 shows the total numbers of acoustic modes whose parametric gain R are greater than 1 as a function of mirror radius curvature for sapphire and fused silica respectively. The optimum tuning of fused silica leads to 2 modes with R of 1.5 and 2.5 when the radius curvature increased to 2.135 km. In sapphire there are no modes with R greater than 1 at the optimum tuning point corresponding to about 2.092 km and 2.127 km radius of curvature.

Tuning the arm cavity radius of curvature also changes the TEM$_{00}$ mode waist size and may mismatch the arm cavity with the recycling cavities. Over modest tuning ranges, this effect is small. For example, when the radius of curvature of the AdvLIGO end test mass changes from 2.076 km to 2.066 km, the arm cavity beam waist changes from 1.15cm to 1.13 cm. The introduced loss due to the mode mismatching is ~300ppm which is acceptable in relation to the recycling mirror transmission (6% for the power recycling mirror and 7% for the signal recycling mirror).

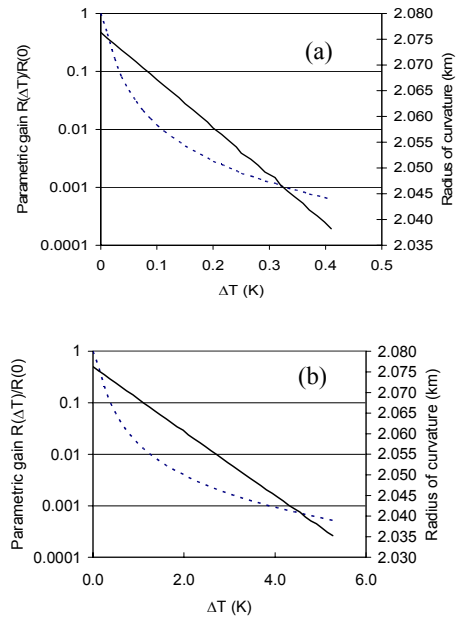

FIG. 2. The dependence of the relative parametric gain R (dotted line) and the mirror radius of curvature (solid line) on the maximum temperature difference across the test mass if considering only one acoustic mode, (a) for sapphire, and (b) for fused silica

In summary, it is inevitable that parametric instabilities will appear in AdvLIGO. By thermally tuning the arm cavity mirror radius of curvature we can tune the cavity away from instability in the case of sapphire test masses or minimise the instability gain in the case of fused silica test masses. Thermal tuning is feasible and need not introduce extra noise. While the data has been applied to AdvLIGO parameters, it is also directly relevant to the VIRGO interferometer.

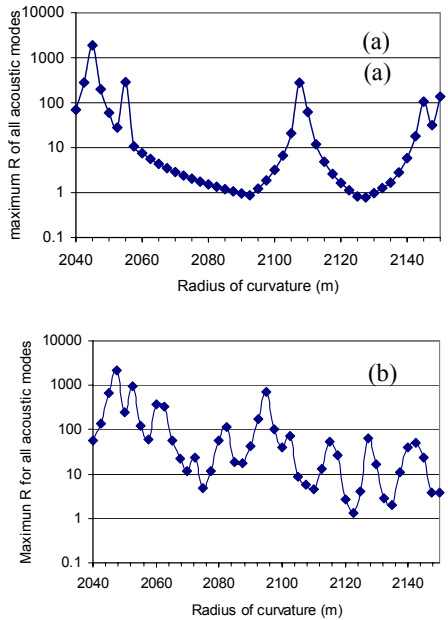

FIG. 3. The maximum parametric gain R of all acoustic modes as a function of mirror radius of curvature, (a) for sapphire and (b) for fused silica.

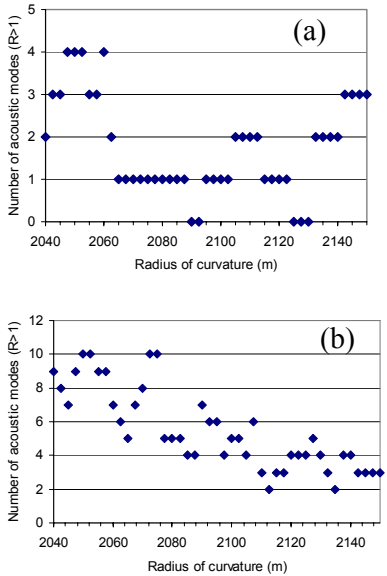

FIG. 4. The numbers of acoustic modes with R>1 when tuning the radius of curvature for (a) sapphire, and (b) fused silica

We note that the instabilities discussed here refer to single cylindrical test masses. Each test mass will experience instability and for non-cylindrical symmetry there will be twice as many acoustic modes .The optimum tuning for suspended pairs of test masses will require further study.

Braginsky et al [15] has proposed the use of small but high finesse detuned cavities as a means of low noise "tranquilizing" of parametric instabilities. The extra cavities needed in the scheme create extra complexity into an already complex system. Feedback schemes similar to the demonstrated cold damping of thermal noise [16] could be another solution, again adding complexity.

A similar analysis for the Gingin High Optical Power Facility shows that this facility is ideally suited for experimentally study of parametric instability. These results will be presented elsewhere

This is a project of Australian Consortium for Gravitational Astronomy (ACIGA), supported by Australian Research Council.

Table III. Sapphire

| $\omega_m/2\pi$ (kHz) | Mode of $\omega_1$ | $\omega_1/2\pi$ (kHz) | $\Lambda_1$ | Mode of $\omega_{1a}$ | $\omega_{1a}/2\pi$ (kHz) | $\Lambda_{1a}$ | R |
|---|---|---|---|---|---|---|---|
| 26.445 | LGM20 | 28.290 | 0.0092 | N/A | 27.630 | 0.0000 | 0.0 |
| 27.399 | LGM20 | 28.290 | 0.1586 | LGM22 | 27.630 | 0.0004 | 0.1 |
| 30.824 | LGM10 | 32.895 | 1.3664 | LGM13 | 32.235 | 0.2336 | -2.6 |
| 30.824 | LGM01 | 28.290 | 0.6072 | LGM03 | 27.630 | 0.0002 | 0.1 |
| 31.070 | HGM10 | 32.895 | 0.5261 | N/A | 32.235 | 0.0000 | 0.1 |
| 33.388 | HGM10 | 32.895 | 0.6888 | N/A | 32.235 | 0.0000 | 1.9 |
| 34.422 | LGM10 | 32.895 | 0.9567 | LGM13 | 32.235 | 0.0392 | 0.1 |
| 63.806 | LGM01 | 65.790 | 1.5307 | LGM03 | 65.130 | 0.4801 | -0.6 |
| 66.285 | LGM01 | 65.790 | 0.0044 | LGM41 | 65.130 | 0.1100 | -0.2 |
| 66.445 | LGM01 | 65.790 | 0.3988 | LGM03 | 65.130 | 0.4933 | -0.4 |
| 66.715 | LGM20 | 65.790 | 0.2562 | LGM22 | 65.130 | 0.3715 | -0.3 |
| 67.105 | LGM10 | 70.395 | 0.0007 | N/A | 69.735 | 0.0000 | 0.0 |
| 67.509 | LGM10 | 70.395 | 0.0001 | N/A | 69.735 | 0.0000 | 0.0 |
| 69.479 | HGM10 | 70.395 | 0.1733 | HGM70 | 69.735 | 0.0182 | -0.3 |
| 69.822 | HGM10 | 70.395 | 0.0002 | N/A | 69.735 | 0.0000 | 0.0 |
| 69.935 | LGM10 | 70.395 | 0.0283 | LGM32 | 69.735 | 0.1322 | -2.2 |
| 70.315 | LGM10 | 70.395 | 0.0006 | N/A | 69.735 | 0.0000 | 0.0 |
| 70.716 | HGM10 | 70.395 | 0.0001 | N/A | 69.735 | 0.0000 | 0.0 |
| 71.046 | HGM10 | 70.395 | 1.1455 | HGM70 | 69.735 | 0.3292 | -0.5 |
| 71.293 | HGM10 | 70.395 | 0.2293 | HGM70 | 69.735 | 0.3280 | -0.6 |
| 71.576 | HGM10 | 70.395 | 0.0005 | N/A | 69.735 | 0.0000 | 0.0 |

#the tables list all acoustic modes in the frequency range with overlap factors for Stokes modes greater than 0.0001. If there is not any anti-Stokes mode which has overlap factors with the particular acoustic mode greater than 0.0001, "N/A" was filled in. The reason is that when overlap factors is less than 0.0001 the R is definitely small than 1.

HGM$_{mn}$ stands for Hermite-Gausian modes of index of *m* and *n*.
LGM$_{mn}$ stands for Laguerre-Gaussian modes of index of *m* and *n*.

Table IV. Fused silica

| $\omega_m/2\pi$ (kHz) | Mode of $\omega_1$ | $\omega_1/2\pi$ (kHz) | $\Lambda_1$ | Mode of $\omega_{1a}$ | $\omega_{1a}/2\pi$ (kHz) | $\Lambda_{1a}$ | R |
|---|---|---|---|---|---|---|---|
| 27.045 | HGM20 | 28.290 | 0.0051 | LGM60 | 27.630 | 0.0036 | -0.1 |
| 27.547 | HGM20 | 28.290 | 0.0356 | HGM06 | 27.630 | 0.0028 | -0.4 |
| 27.918 | HGM20 | 28.290 | 0.0352 | LGM22 | 27.630 | 0.0004 | 0.1 |
| 28.729 | HGM20 | 28.290 | 0.4886 | LGM22 | 27.630 | 0.0330 | 1.0 |
| 29.114 | LGM01 | 28.290 | 0.0003 | LGM41 | 27.630 | 0.0219 | -0.1 |
| 29.933 | LGM01 | 28.290 | 2.0996 | LGM03 | 27.630 | 0.0697 | 0.3 |
| 30.156 | HGM10 | 32.895 | 0.0098 | N/A | 32.235 | 0.0000 | 0.0 |
| 31.247 | LGM10 | 32.895 | 0.0989 | LGM13 | 32.235 | 0.3807 | -4.1 |
| 31.251 | HGM10 | 32.895 | 2.8769 | HGM70 | 32.235 | 0.0020 | 0.4 |
| 31.353 | LGM10 | 32.895 | 0.2270 | LGM32 | 32.235 | 0.1699 | -2.1 |
| 31.554 | HGM10 | 32.895 | 0.1633 | HGM70 | 32.235 | 0.0032 | 0.0 |
| 31.743 | HGM10 | 32.895 | 0.0003 | HGM16 | 32.235 | 0.0035 | -0.1 |
| 32.634 | HGM10 | 32.895 | 0.5242 | HGM70 | 32.235 | 0.0041 | 2.6 |
| 32.816 | LGM10 | 32.895 | 0.1483 | LGM32 | 32.235 | 0.1418 | 5.1 |
| 33.124 | HGM10 | 32.895 | 0.6304 | HGM70 | 32.235 | 0.0061 | 4.0 |
| 33.195 | LGM10 | 32.895 | 0.0002 | LGM51 | 32.235 | 0.0179 | -0.2 |
| 33.354 | HGM10 | 32.895 | 2.2672 | HGM70 | 32.235 | 0.0322 | 3.3 |
| 33.453 | HGM10 | 32.895 | 0.0036 | HGM16 | 32.235 | 0.0157 | -0.1 |
| 33.552 | HGM10 | 32.895 | 0.4634 | HGM70 | 32.235 | 0.0507 | 0.1 |
| 33.615 | HGM10 | 32.895 | 0.0070 | HGM16 | 32.235 | 0.0081 | 0.0 |
| 34.311 | LGM10 | 32.895 | 0.0174 | LGM51 | 32.235 | 0.0043 | 0.0 |
| 34.572 | LGM10 | 32.895 | 0.9214 | LGM13 | 32.235 | 0.4202 | -0.7 |
| 34.767 | LGM10 | 32.895 | 0.0222 | LGM51 | 32.235 | 0.0648 | -0.1 |
| 34.948 | LGM10 | 32.895 | 0.1352 | LGM32 | 32.235 | 0.1500 | -0.2 |
| 64.006 | LGM01 | 65.790 | 0.4227 | LGM03 | 65.130 | 1.2658 | -1.1 |
| 64.108 | LGM01 | 65.790 | 0.2520 | LGM03 | 65.130 | 0.4976 | -0.5 |
| 64.115 | LGM20 | 65.790 | 0.0619 | N/A | 65.130 | 0.0000 | 0.0 |
| 64.349 | LGM01 | 65.790 | 0.0038 | N/A | 65.130 | 0.0000 | 0.0 |
| 64.387 | LGM01 | 65.790 | 0.2766 | LGM41 | 65.130 | 0.0038 | 0.0 |
| 64.521 | LGM20 | 65.790 | 0.0004 | N/A | 65.130 | 0.0000 | 0.0 |
| 64.666 | LGM20 | 65.790 | 0.1049 | N/A | 65.130 | 0.0000 | 0.0 |
| 64.890 | LGM01 | 65.790 | 0.0010 | N/A | 65.130 | 0.0000 | 0.0 |
| 65.098 | LGM01 | 65.790 | 0.3161 | HGM60 | 65.130 | 0.1659 | -4.9 |
| 65.332 | LGM20 | 65.790 | 0.1343 | LGM22 | 65.130 | 0.1511 | -2.1 |
| 65.615 | LGM20 | 65.790 | 0.0070 | N/A | 65.130 | 0.0000 | 0.0 |
| 65.781 | LGM01 | 65.790 | 0.0043 | N/A | 65.130 | 0.0000 | 1.2 |
| 65.872 | LGM20 | 65.790 | 0.0657 | N/A | 65.130 | 0.0000 | 0.8 |
| 65.880 | LGM20 | 65.790 | 0.0797 | N/A | 65.130 | 0.0000 | 0.8 |
| 65.893 | LGM20 | 65.790 | 0.0883 | N/A | 65.130 | 0.0000 | 0.7 |
| 65.954 | LGM20 | 65.790 | 0.0002 | N/A | 65.130 | 0.0000 | 0.0 |
| 66.233 | LGM20 | 65.790 | 0.0002 | N/A | 65.130 | 0.0000 | 0.0 |
| 66.239 | LGM20 | 65.790 | 0.0047 | N/A | 65.130 | 0.0000 | 0.0 |
| 66.298 | LGM01 | 65.790 | 0.0020 | N/A | 65.130 | 0.0000 | 0.0 |
| 66.461 | LGM20 | 65.790 | 0.2048 | LGM22 | 65.130 | 0.2207 | -0.1 |
| 66.463 | LGM20 | 65.790 | 0.4997 | LGM22 | 65.130 | 0.5383 | -0.2 |
| 66.548 | LGM01 | 65.790 | 1.6002 | LGM03 | 65.130 | 1.1915 | -0.4 |
| 66.625 | LGM01 | 65.790 | 0.0002 | N/A | 65.130 | 0.0000 | 0.0 |
| 66.793 | LGM01 | 65.790 | 0.0083 | N/A | 65.130 | 0.0000 | 0.0 |
| 66.875 | LGM20 | 65.790 | 1.0110 | LGM22 | 65.130 | 0.2917 | 0.0 |
| 67.124 | LGM01 | 65.790 | 0.0002 | N/A | 65.130 | 0.0000 | 0.0 |
| 67.208 | LGM01 | 65.790 | 0.0004 | N/A | 65.130 | 0.0000 | 0.0 |
| 67.273 | LGM01 | 65.790 | 1.9113 | N/A | 65.130 | 0.0000 | 0.1 |
| 67.337 | LGM20 | 65.790 | 0.0005 | N/A | 65.130 | 0.0000 | 0.0 |
| 67.379 | LGM20 | 65.790 | 0.0487 | LGM22 | 65.130 | 0.0904 | 0.0 |
| 67.667 | LGM20 | 65.790 | 0.1359 | N/A | 65.130 | 0.0000 | 0.0 |
| 67.701 | LGM01 | 65.790 | 0.0991 | N/A | 65.130 | 0.0000 | 0.0 |
| 67.725 | HGM10 | 70.395 | 0.1081 | HGM70 | 69.735 | 0.3370 | -0.2 |
| 67.840 | LGM20 | 65.790 | 0.0122 | N/A | 65.130 | 0.0000 | 0.0 |
| 67.842 | HGM10 | 70.395 | 0.0001 | N/A | 69.735 | 0.0000 | 0.0 |
| 67.853 | LGM10 | 70.395 | 0.8217 | LGM13 | 69.735 | 0.6087 | -0.4 |
| 67.949 | LGM10 | 70.395 | 0.0003 | N/A | 69.735 | 0.0000 | 0.0 |
| 67.958 | LGM10 | 70.395 | 0.0005 | N/A | 69.735 | 0.0000 | 0.0 |
| 68.029 | HGM10 | 70.395 | 0.0155 | N/A | 69.735 | 0.0000 | 0.0 |
| 68.112 | LGM10 | 70.395 | 0.0112 | N/A | 69.735 | 0.0000 | 0.0 |
| 68.377 | HGM10 | 70.395 | 0.0017 | N/A | 69.735 | 0.0000 | 0.0 |
| 68.443 | HGM10 | 70.395 | 0.0002 | N/A | 69.735 | 0.0000 | 0.0 |
| 68.487 | HGM10 | 70.395 | 0.0017 | N/A | 69.735 | 0.0000 | 0.0 |
| 68.491 | HGM10 | 70.395 | 0.0018 | N/A | 69.735 | 0.0000 | 0.0 |
| 68.797 | LGM10 | 70.395 | 0.0011 | N/A | 69.735 | 0.0000 | 0.0 |
| 68.818 | LGM10 | 70.395 | 0.0005 | N/A | 69.735 | 0.0000 | 0.0 |
| 68.834 | LGM10 | 70.395 | 0.0002 | N/A | 69.735 | 0.0000 | 0.0 |
| 68.862 | LGM10 | 70.395 | 0.0056 | N/A | 69.735 | 0.0000 | 0.0 |
| 68.906 | HGM10 | 70.395 | 0.0409 | N/A | 69.735 | 0.0000 | 0.0 |
| 68.949 | HGM10 | 70.395 | 0.1202 | HGM70 | 69.735 | 0.7778 | -2.4 |
| 69.009 | HGM10 | 70.395 | 0.0001 | N/A | 69.735 | 0.0000 | 0.0 |
| 69.338 | LGM10 | 70.395 | 0.0002 | N/A | 69.735 | 0.0000 | 0.0 |
| 69.496 | HGM10 | 70.395 | 0.0009 | N/A | 69.735 | 0.0000 | 0.0 |
| 69.594 | HGM10 | 70.395 | 0.0001 | N/A | 69.735 | 0.0000 | 0.0 |
| 69.671 | HGM10 | 70.395 | 0.0515 | N/A | 69.735 | 0.0000 | 0.0 |
| 69.677 | HGM10 | 70.395 | 0.0214 | N/A | 69.735 | 0.0000 | 0.0 |
| 69.695 | HGM10 | 70.395 | 0.0038 | N/A | 69.735 | 0.0000 | 0.0 |
| 69.864 | HGM10 | 70.395 | 0.1752 | HGM70 | 69.735 | 0.2944 | -2.6 |
| 69.908 | HGM10 | 70.395 | 0.0056 | N/A | 69.735 | 0.0000 | 0.0 |
| 69.966 | HGM10 | 70.395 | 0.0074 | N/A | 69.735 | 0.0000 | 0.0 |
| 69.978 | LGM10 | 70.395 | 0.0140 | N/A | 69.735 | 0.0000 | 0.0 |
| 70.113 | HGM10 | 70.395 | 0.1308 | HGM70 | 69.735 | 0.2389 | -1.4 |
| 70.203 | HGM10 | 70.395 | 0.0002 | N/A | 69.735 | 0.0000 | 0.0 |
| 70.278 | LGM10 | 70.395 | 0.0000 | N/A | 69.735 | 0.0000 | 0.0 |
| 70.362 | LGM10 | 70.395 | 0.0001 | N/A | 69.735 | 0.0000 | 0.0 |
| 70.400 | HGM10 | 70.395 | 0.0016 | N/A | 69.735 | 0.0000 | 0.5 |
| 70.404 | HGM10 | 70.395 | 0.0014 | N/A | 69.735 | 0.0000 | 0.3 |
| 70.454 | HGM10 | 70.395 | 0.0001 | N/A | 69.735 | 0.0000 | 0.0 |
| 70.533 | HGM10 | 70.395 | 0.0002 | N/A | 69.735 | 0.0000 | 0.0 |
| 70.537 | LGM10 | 70.395 | 0.0004 | N/A | 69.735 | 0.0000 | 0.0 |
| 70.561 | HGM10 | 70.395 | 0.0028 | N/A | 69.735 | 0.0000 | 0.0 |
| 70.761 | HGM10 | 70.395 | 2.9844 | HGM70 | 69.735 | 0.2024 | 1.2 |
| 70.843 | LGM10 | 70.395 | 0.0017 | N/A | 69.735 | 0.0000 | 0.0 |
| 70.970 | LGM10 | 70.395 | 0.0003 | N/A | 69.735 | 0.0000 | 0.0 |
| 70.974 | HGM10 | 70.395 | 0.0228 | N/A | 69.735 | 0.0000 | 0.0 |
| 71.040 | HGM10 | 70.395 | 0.0165 | N/A | 69.735 | 0.0000 | 0.0 |
| 71.109 | HGM10 | 70.395 | 0.0047 | N/A | 69.735 | 0.0000 | 0.0 |
| 71.151 | HGM10 | 70.395 | 0.0005 | N/A | 69.735 | 0.0000 | 0.0 |
| 71.172 | LGM10 | 70.395 | 0.1423 | LGM13 | 69.735 | 0.0284 | 0.0 |
| 71.207 | HGM10 | 70.395 | 0.0572 | N/A | 69.735 | 0.0000 | 0.0 |
| 71.243 | HGM10 | 70.395 | 0.0002 | N/A | 69.735 | 0.0000 | 0.0 |
| 71.262 | HGM10 | 70.395 | 0.3244 | HGM70 | 69.735 | 0.0404 | 0.0 |
| 71.377 | HGM10 | 70.395 | 0.1099 | LGM51 | 69.735 | 0.0003 | 0.0 |
| 71.412 | HGM10 | 70.395 | 0.0002 | N/A | 69.735 | 0.0000 | 0.0 |
| 71.454 | HGM10 | 70.395 | 0.0003 | N/A | 69.735 | 0.0000 | 0.0 |
| 71.486 | HGM10 | 70.395 | 0.0001 | N/A | 69.735 | 0.0000 | 0.0 |
| 71.518 | HGM10 | 70.395 | 0.0043 | N/A | 69.735 | 0.0000 | 0.0 |
| 71.533 | HGM10 | 70.395 | 0.0323 | N/A | 69.735 | 0.0000 | 0.0 |
| 71.584 | HGM10 | 70.395 | 0.0207 | N/A | 69.735 | 0.0000 | 0.0 |
| 71.710 | LGM10 | 70.395 | 0.0015 | N/A | 69.735 | 0.0000 | 0.0 |
| 71.802 | HGM10 | 70.395 | 0.4840 | HGM70 | 69.735 | 0.3877 | -0.2 |
| 72.088 | HGM10 | 70.395 | 0.0002 | N/A | 69.735 | 0.0000 | 0.0 |
| 72.098 | HGM10 | 70.395 | 0.0023 | N/A | 69.735 | 0.0000 | 0.0 |
| 72.249 | LGM10 | 70.395 | 0.0417 | N/A | 69.735 | 0.0000 | 0.0 |
| 72.306 | LGM10 | 70.395 | 0.0337 | N/A | 69.735 | 0.0000 | 0.0 |